\documentstyle[epsfig]{mn}

\newif\ifAMStwofonts

\ifoldfss
  \ifCUPmtlplainloaded \else
    \NewTextAlphabet{textbfit} {cmbxti10} {}
    \NewTextAlphabet{textbfss} {cmssbx10} {}
    \NewMathAlphabet{mathbfit} {cmbxti10} {} 
    \NewMathAlphabet{mathbfss} {cmssbx10} {} 
  \fi
  \ifAMStwofonts
    \ifCUPmtlplainloaded \else
      \NewSymbolFont{upmath} {eurm10}
      \NewSymbolFont{AMSa} {msam10}
      \NewMathSymbol{\upi}     {0}{upmath}{19}
      \NewMathSymbol{\umu}     {0}{upmath}{16}
      \NewMathSymbol{\upartial}{0}{upmath}{40}
      \NewMathSymbol{\leqslant}{3}{AMSa}{36}
      \NewMathSymbol{\geqslant}{3}{AMSa}{3E}

       \let\ge=\geqslant
    \fi
  \fi
\fi 

\ifnfssone
  \newmathalphabet{\mathit}
  \addtoversion{normal}{\mathit}{cmr}{m}{it}
  \addtoversion{bold}{\mathit}{cmr}{bx}{it}
  \newmathalphabet{\mathbfit} 
  \addtoversion{normal}{\mathbfit}{cmr}{bx}{it}
  \addtoversion{bold}{\mathbfit}{cmr}{bx}{it}
  \newmathalphabet{\mathbfss} 
  \addtoversion{normal}{\mathbfss}{cmss}{bx}{n}
  \addtoversion{bold}{\mathbfss}{cmss}{bx}{n}
  \ifAMStwofonts
    \ifCUPmtlplainloaded \else
      \UseAMStwoboldmath
      \makeatletter
      \new@mathgroup\upmath@group
      \define@mathgroup\mv@normal\upmath@group{eur}{m}{n}
      \define@mathgroup\mv@bold\upmath@group{eur}{b}{n}
      \edef\UPM{\hexnumber\upmath@group}
      \new@mathgroup\amsa@group
      \define@mathgroup\mv@normal\amsa@group{msa}{m}{n}
      \define@mathgroup\mv@bold\amsa@group{msa}{m}{n}
      \edef\AMSa{\hexnumber\amsa@group}
      \makeatother
      \mathchardef\upi="0\UPM19
      \mathchardef\umu="0\UPM16
      \mathchardef\upartial="0\UPM40
      \mathchardef\leqslant="3\AMSa36
      \mathchardef\geqslant="3\AMSa3E

       \let\ge=\geqslant
    \fi
  \fi
\fi 

\ifnfsstwo
  \DeclareMathAlphabet{\mathbfit}{OT1}{cmr}{bx}{it}
  \SetMathAlphabet\mathbfit{bold}{OT1}{cmr}{bx}{it}
  \DeclareMathAlphabet{\mathbfss}{OT1}{cmss}{bx}{n}
  \SetMathAlphabet\mathbfss{bold}{OT1}{cmss}{bx}{n}
  \ifAMStwofonts
    \ifCUPmtlplainloaded \else
      \DeclareSymbolFont{UPM}{U}{eur}{m}{n}
      \SetSymbolFont{UPM}{bold}{U}{eur}{b}{n}
      \DeclareSymbolFont{AMSa}{U}{msa}{m}{n}
      \DeclareMathSymbol{\upi}{0}{UPM}{"19}
      \DeclareMathSymbol{\umu}{0}{UPM}{"16}
      \DeclareMathSymbol{\upartial}{0}{UPM}{"40}
      \DeclareMathSymbol{\leqslant}{3}{AMSa}{"36}
      \DeclareMathSymbol{\geqslant}{3}{AMSa}{"3E}

       \let\ge=\geqslant
    \fi
  \fi
\fi 

\ifCUPmtlplainloaded \else
  \ifAMStwofonts \else 
    \def\upi{\pi}
    \def\umu{\mu}
    \def\upartial{\partial}
  \fi
\fi

\title[NGC 7331 dust ring]{SCUBA imaging of NGC 7331 dust 
ring\thanks{Based on observations at the James Clerk Maxwell Telescope.
JCMT is operated by The 
Joint Astronomy Centre on behalf of the Particle Physics and 
Astronomy Research Council of the United Kingdom, the Netherlands 
Organisation for Scientific Research, and the National Research 
Council of Canada.}}
\author[S. Bianchi et al.]
       {S. Bianchi,$^1$
	P.~B. Alton,$^1$
	J.~I.~Davies $^1$
	and M.~Trewhella$^2$\\ 
$^1$Department of Physics and Astronomy, University of Wales 
Cardiff, P.O. Box 913, Cardiff, CF2 3YB, UK\\
$^2$IPAC, M.S. 100-22 Caltech, Pasadena, CA91125, USA}

\begin{document}

\maketitle
\begin{abstract}
We present observations of the spiral galaxy NGC 7331 using the 
Sub-millimetre Common User Bolometer Array (SCUBA) on the James Clark
Maxwell Telescope. We have detected a dust ring of 45 arcsec radius 
(3.3 kpc) at wavelengths of 450 and 850-$\mu$m. The dust ring is in good
correspondence with other observations of the ring in the mid-infrared (MIR), 
CO and radio-continuum, suggesting that the observed dust is associated with
the molecular gas and star formation. A  B-K colour map shows an analogous 
ring structure with an asymmetry about the major axis, consistent with
the extinction being produced by a dust ring.
The derived temperature of the dust lies between 16 and 31 K and the 
gas-to-dust ratio between 150 and 570, depending on the assumed 
dust emission efficiency index ($\beta=$1.5 or 2.). 
\end{abstract}

\section{Introduction}
Modelling of \textit{COBE}-DIRBE observation of our Galaxy has shown 
that dust at $T<$22 K associated with neutral and molecular gas is 
responsible for most of the emission (55-85 per cent) in the far-infrared 
(FIR)~\cite{so94,so97}. Reach et al.~\shortcite{re95},
analysing \textit{COBE}-FIRAS spectral data, find that Galactic emission 
between 100-$\mu$m and 300-$\mu$m is predominantly from dust at $T$=16-23 K.
This range of temperatures is in agreement with theoretical 
calculations for classical grains heated by the interstellar radiation 
field (ISRF)~\cite{di89}. 

In external galaxies, dust temperatures have been mainly measured
using \textit{IRAS} fluxes at 60-$\mu$m and 100-$\mu$m. 
\textit{IRAS} can only trace warm dust at $T\ga$ 30 K~\cite{de90}. 
This is mainly associated with star forming regions, as illustrated 
by the strong correlation between H$\alpha$ and \textit{IRAS}-FIR 
emission~\cite{de95}.
Gas-to-dust ratios derived from \textit{IRAS} are a factor of 10 larger than 
the Galactic value~\cite{de90}. Assuming that the Galaxy is typical of
spirals, the discrepancy can be explained by a colder dust component
($T\approx$ 15 K) that contributes to the bulk of the mass, without 
adding a lot of flux in the window observed by \textit{IRAS}~\cite{de90}.
This is because of the strong dependence of emission on temperature
($\propto T^{4+\beta}, \beta=1-2$).

Estimates of the temperature of the main dust component depend on the 
availability of data at $\lambda>$100-$\mu$m.  Recently, Alton et 
al.~\shortcite{al98} have measured a $\langle T\rangle \approx$ 20 K 
for a sample of 7 spiral galaxies, using 
\textit{ISO}PHOT 200-$\mu$m maps combined with HiRes \textit{IRAS} 
maps at 100-$\mu$m, 10K lower than the temperatures measured 
using \textit{IRAS} HiRes 60-$\mu$m data.
The mean gas-to-dust ratio is then $\approx 230$, closer to the canonical 
Galactic value of $160$~\cite{so94}.
Newly available FIR maps, despite their low resolution 
(117 arcsec for \textit{ISO}PHOT data) have revealed that the spatial 
distribution of dust is more extended than that of the \textit{IRAS} 
and K-band (stellar) emission~\cite{al98,da98}.

In this {\em letter} we describe sub-millimetre (submm) observations of 
NGC 7331 with SCUBA. Maps produced by 
SCUBA have a far higher resolution than \textit{ISO}PHOT and 
HiRes \textit{IRAS} long wavelength observations and give us detailed
spatial information about the extent of cold dust.
NGC 7331 is an Sb galaxy~\cite{rc3} at a distance of 15.1 Mpc~\cite{hu98},
giving an apparent scale of 73 pc arcsec$^{-1}$. 

NGC 7331 shows a ring structure in CO emission~\cite{to97,vl96,yo82},
in radio-continuum~\cite{co94} and in MIR images from
\textit{ISO}CAM (6.75-$\mu$m$<\lambda<$15-$\mu$m)~\cite{sm98}.
Smith~\shortcite{sm98} also shows that these observations are in 
good correspondence with each other, indicating the presence of a
molecular and dust ring of radius $\approx$ 45 arcsec (3kpc). There is 
also a good match with H$\alpha$+[N\textsc{II}] features~\cite{po89}. 
The central region (radius $\approx$2 arcmin) is partially depleted of 
neutral hydrogen~\cite{bo81} whilst the molecular gas contributes to 70 per 
cent of the total gas mass inside the optical radius~\cite{de90}. 
A major axis profile at 100-$\mu$m obtained with the Kuiper Airborne
Observatory (KAO)~\cite{sm96} shows a flat-topped FIR distribution
($FWHM \approx$ 35 arcsec), but very little structure can be seen in 
the HiRes-IRAS maps or in the 200-$\mu$m \textit{ISO}PHOT observations 
of Alton et al.~\shortcite{al98} ($FWHM\ga$ 50 arcsec).

The ring may be the result of long-term dynamical 
evolution driven by a bar instability, as proposed by von Linden et
al.~\shortcite{vl96}. Tosaki \& Shioya~\shortcite{to97} relate the ring to
the post starburst status of the galaxy: the central gas is 
consumed in the long evolution after the starburst event.

\section{Observation and data reduction}
NGC 7331 was observed at JCMT with SCUBA on 1997 October 20, 22, 24. 
SCUBA consists of two bolometer arrays able to simultaneously image 
a region of sky of about 2.3 arcmin in diameter: the {\em short-wavelength 
array}, optimised for observing at 450-$\mu$m (91 elements) and the 
{\em long-wavelength array} optimised at 850-$\mu$m (37 elements).
To fully sample the field of view with both the arrays, the 
secondary mirror is moved in a 64-point jiggle pattern. 
For each jiggle position the integration time is 1 s.
Simultaneously, the secondary mirror is also chopping with a frequency 
of 7 Hz to remove the sky background. Every 16 steps of the jiggle pattern, 
the telescope nods, to remove slowly varying atmospheric gradients.
We used a chop throw of 180 arcsec perpendicular to the major axis of NGC 
7331.

We frequently determined the transparency of the atmosphere during
each night by measuring sky emission at several elevations. Sky conditions
were stable during most of the observing run with zenith 
optical depths of $\tau_{450}=0.6-0.7$ and $\tau_{850}=0.13-0.15$.
During the first night atmospheric opacity was higher, with $\tau_{450}>3$ 
and $\tau_{850}=0.5$. The telescope pointing was checked every hour against a 
bright point source close to our target: {\em rms} pointing errors
were $\approx$ 3 arcsec in both azimuth and elevation.  Several images 
of Uranus were taken for photometric calibration.

Data reduction was carried out using the \textsc{STARLINK} package 
\textsc{SURF}~\cite{sur,coo}. After subtracting the {\em off-source} 
signal, images were flat-fielded and corrected for atmospheric extinction. 
Noisy bolometers were masked. 
Each image was corrected for systematic noise variations using stable 
bolometers that appeared free of source emission. Spikes from transient 
detections were removed by applying a $3\sigma$ clip. 

A calibration constant for each night was computed from the Uranus maps.
Comparing data for each night we derived a relative error in calibration  
of 14 per cent and 7 per cent, for 450-$\mu$m and 850-$\mu$m respectively.  
Uranus images were also used to measure the beam size and the contribution 
of side lobes. 
The measured HPBW is 10 arcsec at 450-$\mu$m and 15 arcsec for 850-$\mu$m. 
The side lobes can introduce quite large systematic errors: for example, 
20 per cent of the emission of a point source goes in the side lobes
for the long wavelength and 60 per cent for the short wavelength.
In the following, values of integrated flux are corrected for the side-lobe 
pickup.

\begin{figure*}
\centerline{
\epsfig{figure=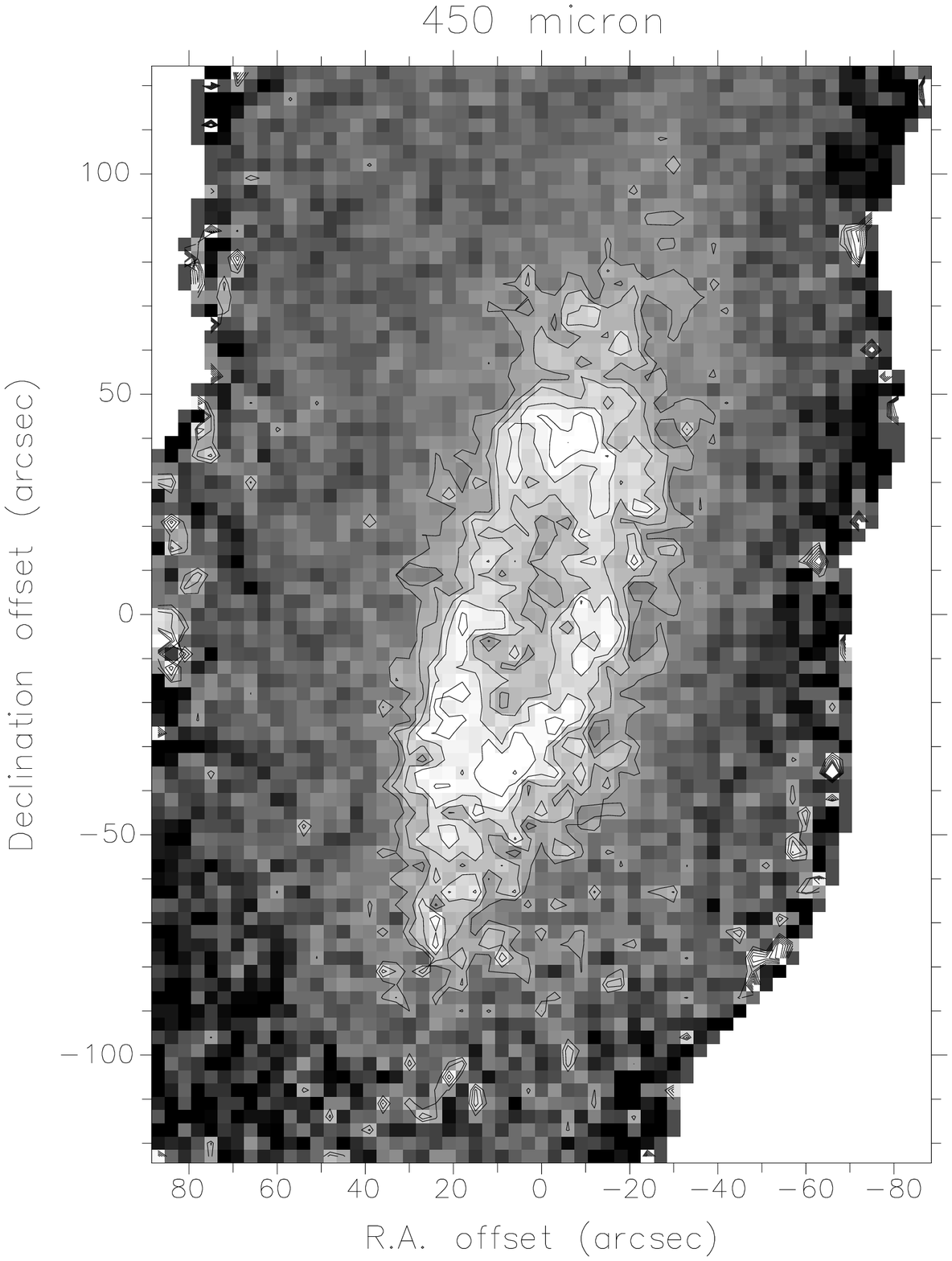,width=6.5cm,bbllx=0pt,bburx=520pt,bblly=80pt,bbury=750pt}
\epsfig{figure=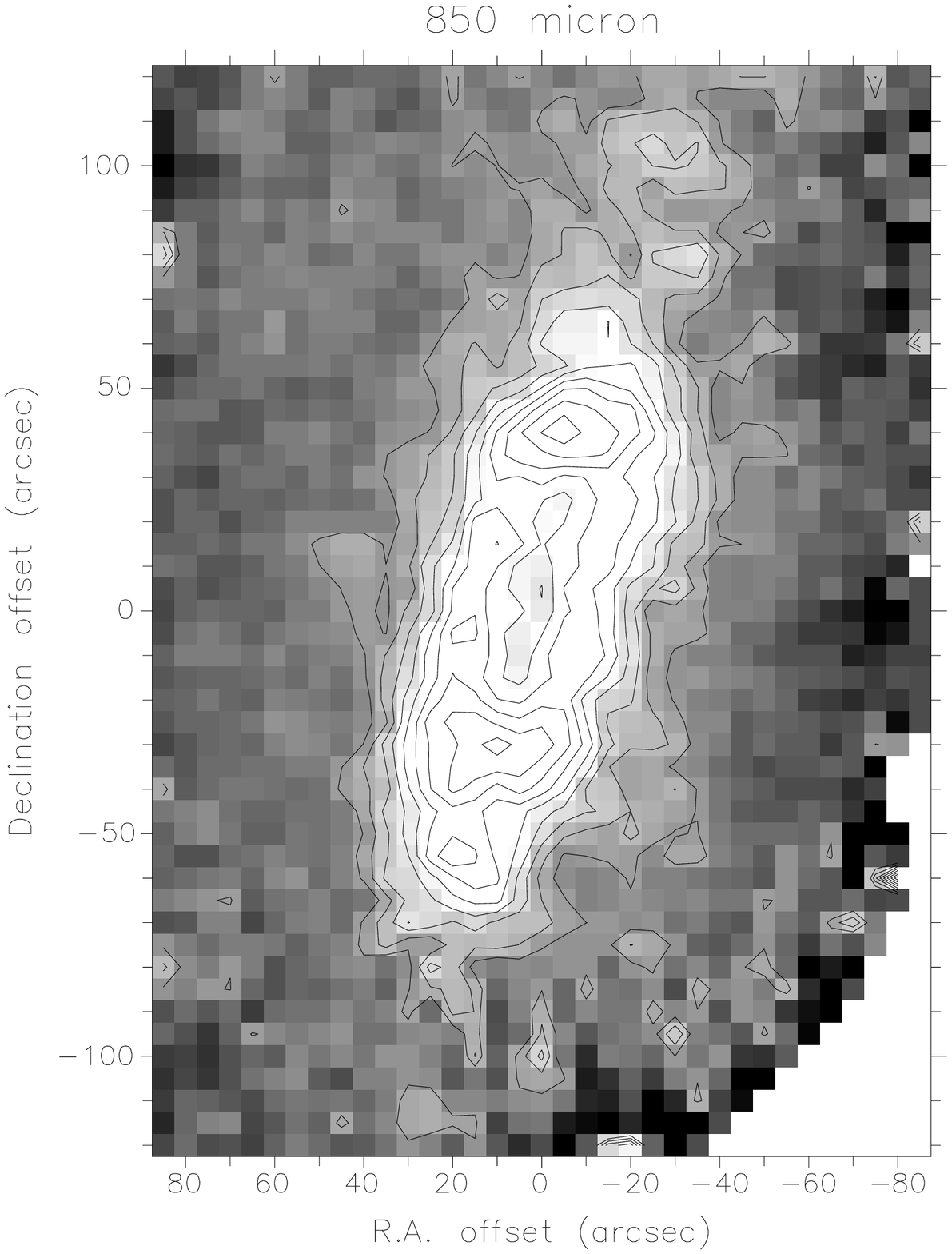,width=6.5cm,bbllx=0pt,bburx=520pt,bblly=80pt,bbury=750pt}
}
\centerline{
\epsfig{figure=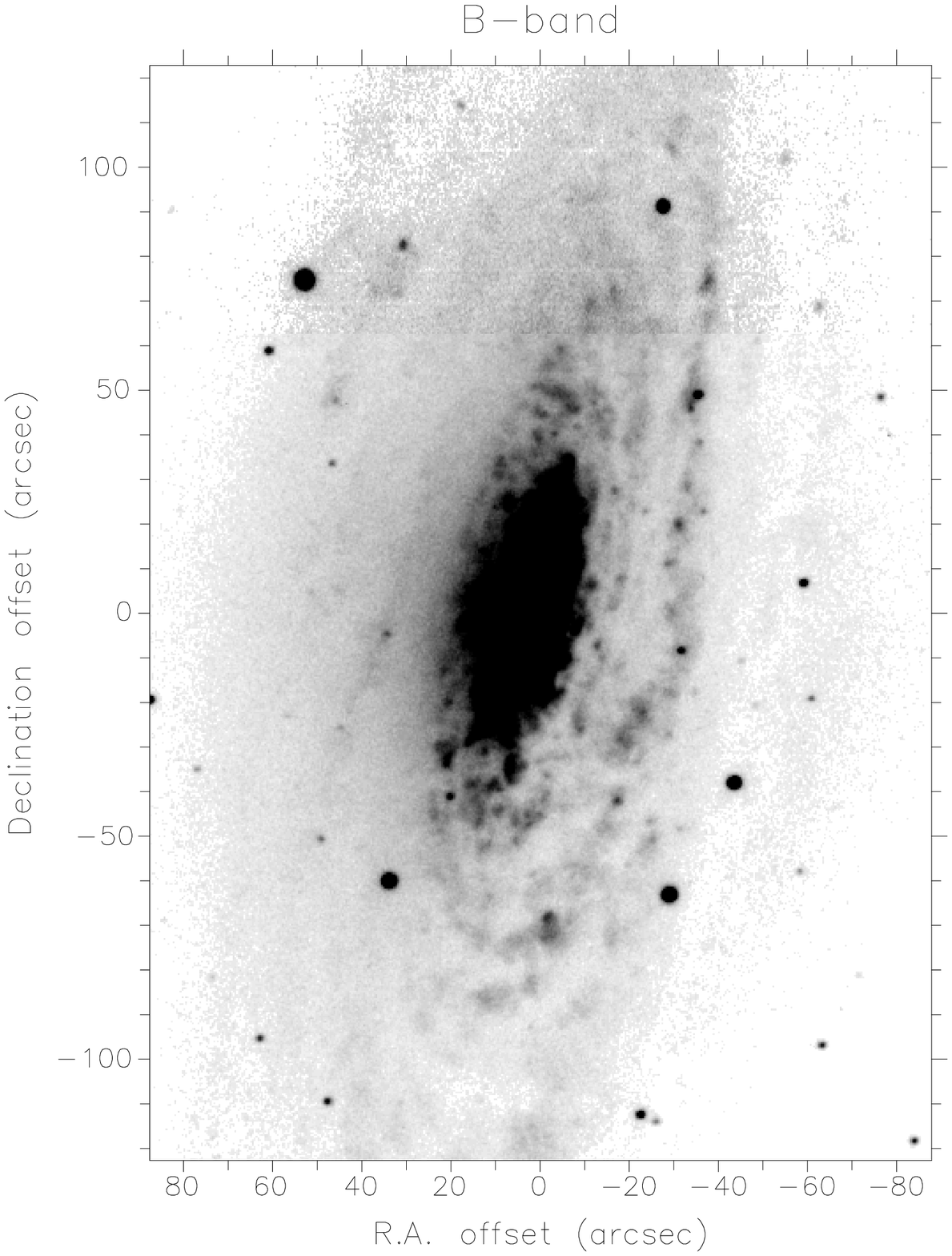,width=6.5cm,bbllx=0pt,bburx=520pt,bblly=80pt,bbury=750pt}
\epsfig{figure=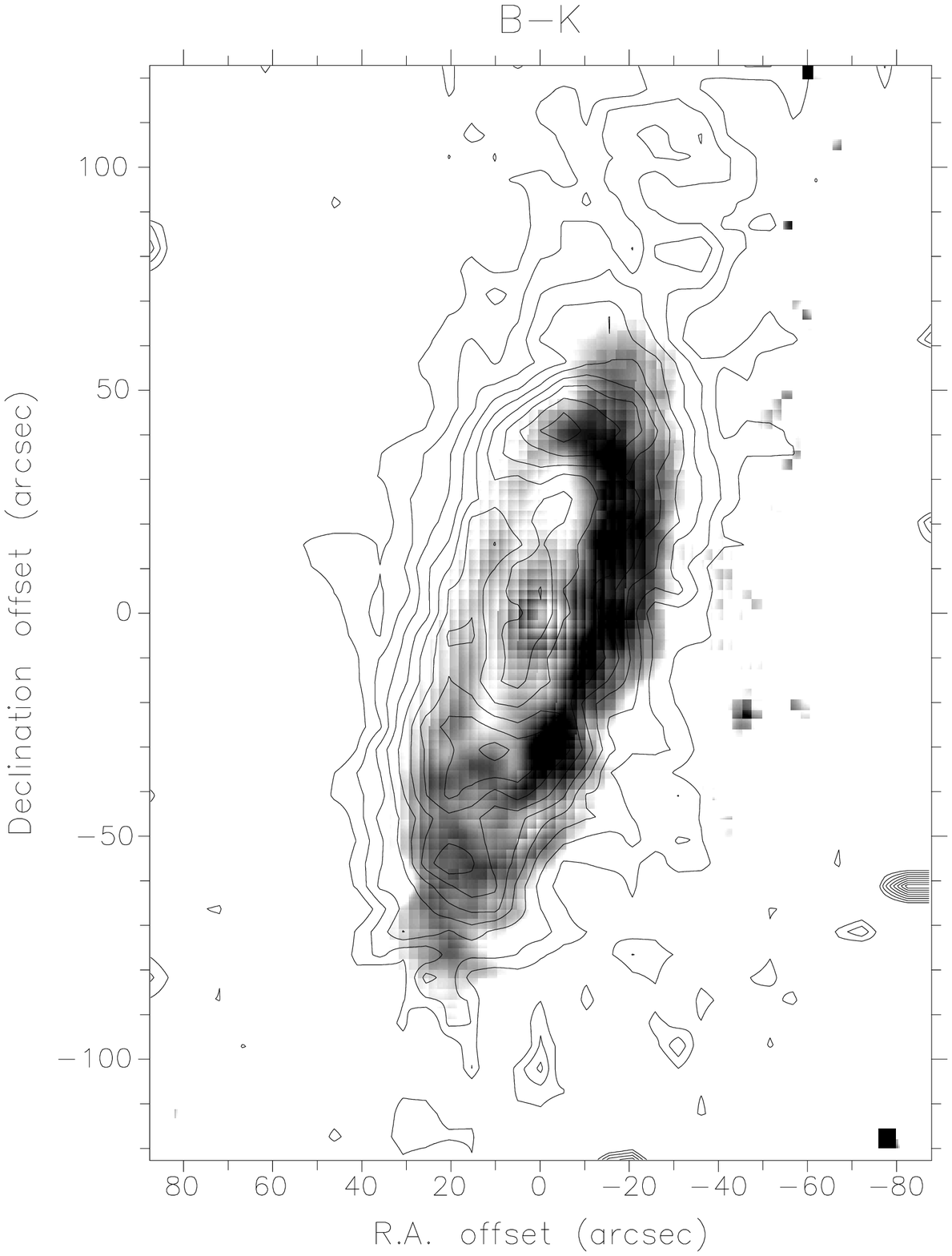,width=6.5cm,bbllx=0pt,bburx=520pt,bblly=80pt,bbury=750pt}
}
\caption{450-$\mu$m (top-left), 850-$\mu$m  (top-right) B-band (bottom-left) 
and B-K (bottom-right) maps of NGC7331. For submm images, contours 
are overlayed at integer values of the S/N ratio, starting from S/N=3
(for the 450-$\mu$m image this corresponds to isophotes every 80 mJy
beam$^{-1}$ starting from 240 mJy beam$^{-1}$; for 850-$\mu$m, isophotes
every 8 mJy beam$^{-1}$ starting from 24  mJy beam$^{-1}$).  
Contours at 850-$\mu$m are also overlayed on the B-K image.
}
\label{fig:all}
\end{figure*}

A total of 15 images at 450-$\mu$m and 19 at 850-$\mu$m  were 
re-sampled to an equatorial reference frame, using pixels of 3 
and 5 arcsec, respectively. The final map covers an area of 3.3 
arcmin x 5.8 arcmin.  The total {\em on-source} integration 
time for the central part of the image is 5700s and 8300s.
Maps are shown in Fig.~\ref{fig:all}; contours with
$S/N \ge 3$ are over-plotted. The short wavelength image clearly
shows a poorer signal to noise, mainly due to the higher 
emissivity of the sky at  450-$\mu$m compared to 850-$\mu$m: 
the sky noise is 80 mJy beam$^{-1}$ at  450-$\mu$m and 8 
mJy beam$^{-1}$ at 850-$\mu$m.

In this work we have also used B and K-band observations. Images were 
obtained from the Skinakas observatory Crete (B) and at the Wyoming Infrared 
Observatory (K). The optical and NIR data were reduced and calibrated in the 
standard way~\cite{tr97}. The PSF has a FWHM=1.5 arcsec for the B-band image 
and 2.5 arcsec for the K-band.
Optical images were aligned in a RA/Dec frame using field stars in 
the HST guide star catalogue. The peak emission in B-band image occurs
at RA=22$^h$ 37$^m$ 04$^s$, Dec=34$^\circ$ 24$'$ 55\fs5 (J2000); 
these coordinates are within 1.5 arcsec of the optical and radio emission 
centres given by Cowan et al.~\shortcite{co94}.
SCUBA images were aligned by assuming the B peak as the centre of 
the frame. Images in B-band and B-K colour are shown in Fig.~\ref{fig:all}.
The B-K colour image has a photometric accuracy of 0.2 mag.

\section{Results \& Discussion}

Our 850-$\mu$m image (Fig.~\ref{fig:all}, top-right) clearly shows a ring 
structure, of $\approx$ 90 arcsec x 30 arcsec (radius$\approx$ 45
arcsec, corresponding to 3.3 kpc). This submm ring matches MIR 
observations, and therefore CO, radio-continuum and  H$\alpha$+[NII], 
as shown in Smith~\shortcite{sm98}.

The brightest parts of the ring are on the north and south sides, where 
intensity peaks at almost the same value (95 mJy beam$^{-1}$, S/N$\approx$ 12).
The east and west side of the ring are less bright and appear clumpy.
As suggested by von Linden et al.~\shortcite{vl96} for the 
CO ring, this morphology may well be due to the inclined view.
There are also two smaller structures attached to the north and south 
sources (seen in our image at positions [5,60] and [20,-50], respectively).
The south one, the brightest, is associated with the origin of
one of the spiral arms in our optical image (Fig.~\ref{fig:all}, bottom-left). 
There is no evidence for a central source, contrary to the MIR 
images~\cite{sm98}.
The ring is also visible in the 450-$\mu$m image (Fig.~\ref{fig:all}, 
top-left) despite the poorer signal. When smoothed to the long
wavelength resolution, the same structures can be seen.

\begin{figure}
\centerline{\epsfig{figure=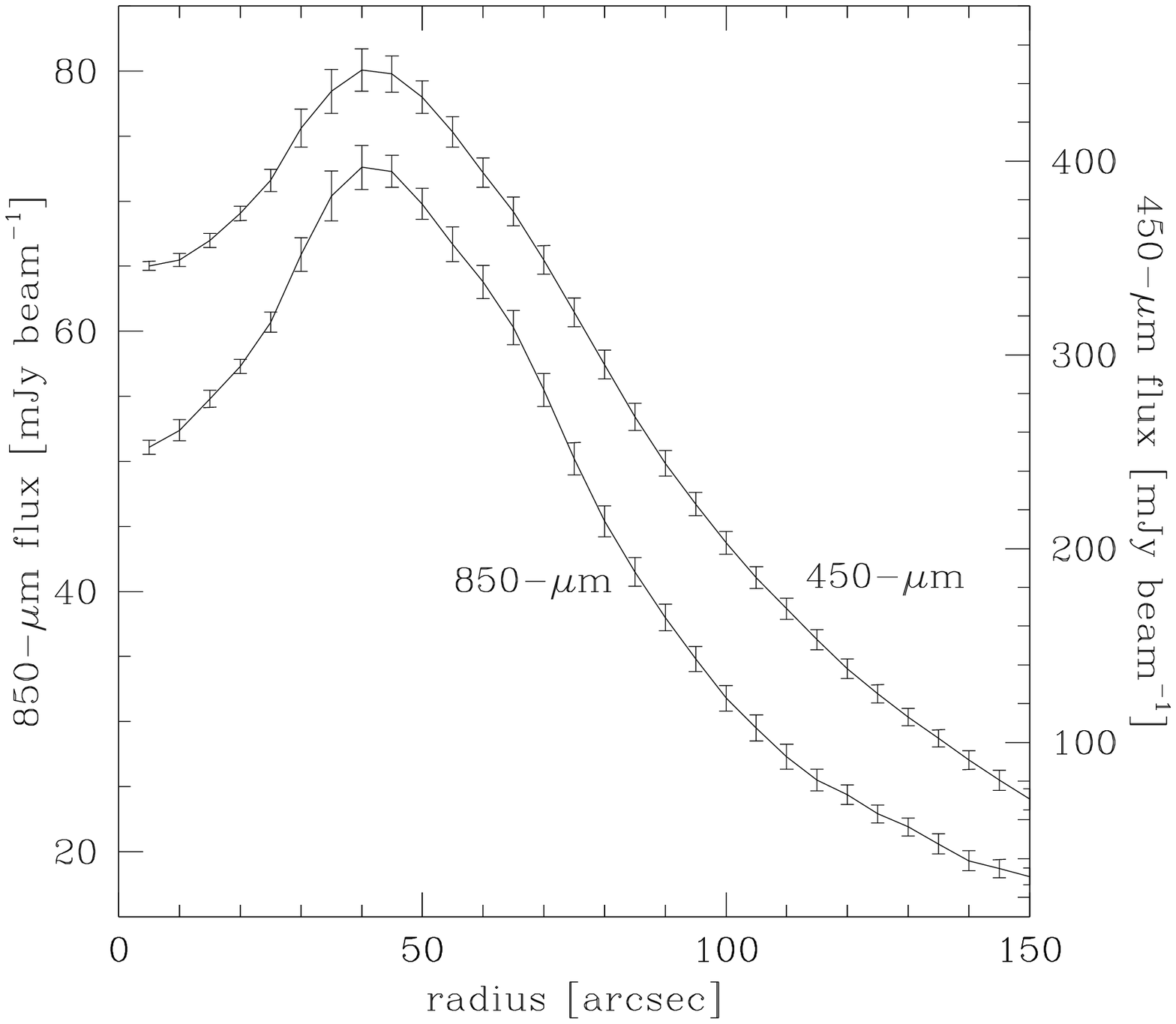,width=8.5cm}}
\caption{Elliptically averaged surface brightness profiles for 850-$\mu
m$ and 450-$\mu m$. Isophote geometrical parameters are described in the 
text. Error bars 
rapresent the standard deviation of the mean inside each elliptical 
annulus.}
\label{fig:profile}
\end{figure}

Radial surface-brightness profiles were produced for the 850-$\mu$m 
and the smoothed 450-$\mu$m images, averaging over elliptic annuli. 
We have adopted a disk 
inclination of $74^\circ$ and $PA=$167$^\circ$~\cite{gg91}. The two
profiles can be followed out to 150 arcsec from the centre (both images
with S/N $>1$), and they appear similar (Fig.~\ref{fig:profile}). 

Using this last aperture, we have computed the integrated fluxes: 
we obtain a flux of 13.4 Jy for 450-$\mu$m and 1.9 Jy for 850-$\mu$m. 
The ratio between the fluxes is therefore $7.0\pm 1.0$, where the error 
is mostly due to the uncertainties in the calibration. 
From this ratio the average dust temperature can be computed, if the 
dependence of dust emissivity $Q_{em}$ on wavelength $\lambda$ is known. 
It is generally assumed that $Q_{em}\propto\lambda^{-\beta}$, with 
$\beta$ increasing from 1 to 2 going from the MIR to the FIR and 
submm~\cite{hi83}. Fits of the Galactic FIR-submm 
spectrum~\cite{re95} suggest a value $\beta=2$ for $\lambda >$200-$\mu$m. 
Millimetre/submm observations of the Galactic plane and cirrus
\cite{ma95} emissions are better described by $\beta=1.5$.
Using $\beta=2$ we derive $T=$16$\pm$3 K ($T=$31$\pm$10 K for $\beta=1.5$).  
The temperature obtained using $\beta=2$ agrees very well with values
for high latitude dust emission in the Galaxy~\cite{re95}.
It is interesting to note that a grey-body model with $\beta=1$ cannot be 
fitted to the data. This is also confirmed by SCUBA observations of 
NGC891~\cite{n891}.

The dust temperature was also derived using the KAO
100-$\mu$m observations presented by Smith \& Harvey~\shortcite{sm96}.
The image at 850-$\mu$m was smoothed to the KAO resolution
(31 arcsec x 41 arcsec) and a profile along the major axis was produced,
with the same orientation as the one shown in their Fig. 5. 
In the central 100 arcsec, where the KAO profile appears flat,
the average flux density per bolometer at 100$\mu$m is 29 Jy.
In the same region the average value for the 850-$\mu$m flux is
60 mJy beam$^{-1}$. From the ratio we derived $T=$23K$\pm$2 for $\beta=2$
($T=29$K$\pm$3 for $\beta=1.5$). 
Alton et al.~\shortcite{al98} fluxes at 100-$\mu$m and \textit{ISO}
200-$\mu$m  give values $T=$17K for $\beta=2$ and 
$T=$19K for $\beta=1.5$.

The correspondence between the submm ring and other observations
suggests that the dust emission is associated with the molecular gas.
Indeed, the values of dust temperature we have presented here are consistent 
with those of the dust component associated with the molecular gas in 
the Galaxy~\cite{so97}.

The B-K map is shown in Fig.~\ref{fig:all}, bottom-right, with the
850-$\mu$m contour superimposed. A ring structure can be seen in the 
colour image, in excellent correspondence with the submm image. 
The ring is delimited on the east and the west side by regions with a redder 
colour with respect to the central one. The colour on the west side is also 
redder than on the east. The asymmetry and the central gap, first observed 
in NIR colors by Telesco, Gatley \& Steward~\shortcite{te82}, can be 
explained with a bulge obscured by an inclined dust ring: on the west side, 
the closer one, 
most of the bulge is behind the dust ring so that bulge light is more 
extinguished and therefore redder than on the east side, the far one. 
Because of the relative lack of dust in the middle of the
structure, bulge light from the central region appears bluer.

\begin{figure}
\centerline{\epsfig{figure=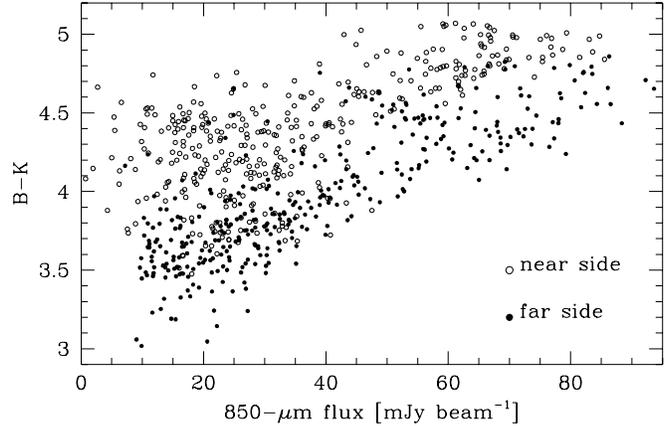,height=6.0cm}}
\caption{Correlation between pixels in a B-K smoothed image and 850-$\mu$m
image, inside a 100 arcsec x 50 arcsec region centred on the galaxy.
Filled circles refer to the far side of the galaxy (i.e. east side)
and open circles to the near side (i.e. west side). The ring structure evident 
in 850-$\mu$m image (Fig.~\ref{fig:all}) correspond to fluxes
bigger than 60 mJy beam$^{-1}$. Each data point has a random error 
of $\approx$0.05 mag.}\label{fig:confro}
\end{figure}

To compare B-K with 850-$\mu$m emission, the colour image was smoothed
to a resolution of 15 arcsec and rebinned on 5 arcsec pixels. We have then 
plotted B-K vs 850-$\mu$m for each pixel in an elliptic region of 
$\approx$ 100 arcsec x 50 arcsec (Fig~\ref{fig:confro}). 
A clear trend is evident: for the same 
850-$\mu$m flux, the closer part of the galaxy is systematically redder than 
the far one, by 0.5 mag. This is true both for the regions with high
flux (i.e. high dust column density) that define the ring and 
for regions external to the ring.  Bulge light extinguished by
a more extended dust disk may be the reason for the colour asymmetry of 
these outer regions. As a check, we repeated the correlation using
pixels of 15 arcsec, but the conclusions are the same: our results are thus
independent of a possible misalignment of the colour and submm images,
which we estimate to be at most 2.5 arcsec.

B-K is more sensitive than optical colours to the presence of 
dust~\cite{bl94}, because of the high ratio between extinction 
in B and K-band ($\approx 14$, using a Galactic extinction law~\cite{whbo}).
Even so, deriving the dust content from the colour image is quite difficult: 
first an {\em a priori} knowledge of the {\em intrinsic} colour distribution
is required, to quantify the {\em reddening}; second, the relative
distribution of dust and stars must be known, to infer, by means of a radiative
transfer model, the relation between the reddening and the optical depth 
(or the dust column density). These two steps are obviously
coupled together~(See for example Xilouris et al. 1997, 1998).

We can easily obtain a lower limit to the optical depth of the ring.
We assume the thickness of the dust structure to be negligible so that 
dust acts like an internal {\em screen} for light emitted by a 
spheroidal bulge. On the near part of the galaxy we assume that all the 
light is extinguished (i.e. the bulge is all behind the screen). 
On the far side the situation is reversed, with the bulge un-extinguished. 
These assumptions are justified by the high inclination of NGC7331 and by 
the luminous bulge, that has an effective radius similar to the ring (31
arcsec; Boroson~1981). Within this model, B-K on the far side
can be regarded as the intrinsic colour of the bulge, and the difference
between the two half as the colour excess due to dust extinction.
From the difference in B-K of 0.5 mag we can thus derive a
optical depth in the V-band of $\approx 0.4$, if a galactic extinction law
is assumed. In reality the dust distribution would have a finite
thickness; besides scattering and clumpiness should be considered.
Inclusion of these features would go in the direction of reducing the
extinction for a given quantity of dust~\cite{bo96,wi96}.

Submm images permit us to derive the V-band optical depth 
of the dust distribution in a way independent of geometry.  
The optical depth $\tau_V$ can be related to the flux at 
$\lambda=$850-$\mu$m using the formula~\cite{hi83}
\[
\tau_V=\frac{1}{2}\left(\frac{Q_{UV}}{Q_{\lambda_0}}\right)
\left(\frac{\lambda}{\lambda_0}\right)^\beta 
\frac{I_{\lambda}}{B_{\lambda}(T)}
\]
where $T$ is the temperature of the dust, $I_{\lambda}$ is the 
intensity (i.e. the flux per beam divided
by the integrated beam size), $B_{\lambda}(T)$ is the Planck 
function and $\beta$ the emissivity index. $Q_{UV}$ is the extinction
efficiency in the ultraviolet range (1500-3000\AA), while $Q_{\lambda_0}$ 
is the emission efficiency at a reference wavelength $\lambda_0$.
The ratio $Q_{UV}/Q_{\lambda_0}$ is quite uncertain: but from 
Hildebrand~\shortcite{hi83} and from Casey~\shortcite{ca91} we have
derived a mean value $Q_{UV}/Q_{\lambda_0}\approx 1200$ at 
$\lambda_0=125\mu m$ for $\beta =$ 2. The value of $Q_{UV}/Q_{\lambda_0}$ for
$\beta=1.5$ has been roughly derived from data in Casey~\shortcite{ca91}: 
we use $Q_{UV}/Q_{\lambda_0}=2100$ at $\lambda_0=125\mu m$. 
Typically, values of $Q_{UV}/Q_{\lambda_0}$ have uncertainties of a factor 
of 2-3.
The factor $1/2$ comes from assuming $\tau_{UV}/\tau_V=2$~\cite{ca91}.
For the ring (mean flux of 70 Jy beam$^{-1}$) we obtain face-on optical
depths of $\tau_V=2.4$ and $9.3$, for $\beta=1.5$ and $2$, respectively. 
At the $1\sigma$ isophote $\tau_V=0.4$ and $0.9$. 
Uncertainties in $\tau_V$ are quite large, because of both the
uncertainties in T and in the value $Q_{UV}/Q_{\lambda_0}$. 
Using temperatures derived with the help of KAO data, optical depths
are $\tau_V=2.6$ and $5.4$, for $\beta=1.5$ and $2$, respectively.
Telesco et al.~\shortcite{te82} derived a $\tau_V\approx$ 3
for the reddest part of the galaxy.

The mass column density of dust can be derived from the optical depth. 
Adopting a mean dust grain radius of 0.1-$\mu$m, a density of $3$ g
cm$^{-3}$ and an extinction efficiency in the V-band 
of 1.5~\cite{hi83,ca91}, we have obtained ring column densities
$N_d=0.3$ M$_\odot $pc$^{-2}$ and $N_d=1.1 $ M$_\odot$ pc$^{-2}$ for 
$\beta=1.5$ and $2.$, respectively ($N_d=0.04$ M$_\odot $pc$^{-2}$ and 
$N_d=0.1$ M$_\odot$ pc$^{-2}$ at $1\sigma$ level).
From CO maps of the east and west side of the ring, Tosaki \& 
Shioya~\shortcite{to97} have deduced a mean mass column density for 
the molecular gas  of 150 M$_\odot $pc$^{-2}$ (we note here that there 
are possible uncertainties related to the smaller beam size of their
observations). An atomic gas column 
density of 20 M$_\odot $pc$^{-2}$ has been derived from 
Bosma~\shortcite{bo81} for the position of the ring. 
The total gas column density for the ring is thus
$N_g=170 $M$_\odot $pc$^{-2}$. The gas-to-dust mass ratio 
is then $150$ for $\beta = 2$ ($570$ for $\beta = 1.5$). 
The value derived for $\beta=2$ is
much closer to the galactic value than \textit{IRAS} based
ratios~\cite{de90} and in agreement with \textit{ISO} based 
determination~\cite{al98,da98}, as well as with optical-NIR radiative
transfer models~(Xilouris et al. 1997, 1998). 

The dust mass inside 150 arcsec is $M_d=4\times 10^7$ M$_\odot$ for 
$\beta=2$ ($M_d=$1 $\times$ 10$^7$ M$_\odot$ for $\beta=1.5$). From
Alton et al.~\shortcite{al98} data, covering a larger area ($\approx$
10 arcmin), we can derive a dust mass of $M_d=$ 1.2 $\times$ 10$^8$ 
M$_\odot$ (for both values of $\beta$). Therefore the total mass is at 
least three times larger than the one we have derived here. 
This shows that there is a large and diffuse component of dust 
($>70$ per cent in mass) that is not associated with the molecular ring.
It is interesting to note that an extended dust distribution may be
the reason why some millimetre and submm observations~\cite{ea89,cl93}
have failed to detect cold dust, because of the small size of 
of the bolometers used compared to the extent of the dust~\cite{cl93}.

\vskip 0.7cm
\noindent {\em Acknowledgements.} We are grateful to Iain Coulson
at JCMT for his help and support during observations and data reduction.
It is also a pleasure to thank Beverly Smith for providing us with KAO 
data, Manolis Xilouris for the B-band image and Harley Thronson for help 
with the K-band data.

\end{document}